
\def\beginsection#1{\vskip .25in\goodbreak\noindent
		    {\bf{#1}}\nobreak
		    \smallskip\nobreak\noindent\hskip -.0mm}

\magnification 1200
\baselineskip=1.2\baselineskip

\def\al{\alpha}
\def\be{\beta}
\def\ga{\gamma}
         
\def\ep{\epsilon}
\def\ze{\zeta}

\def\th{\theta}
\def\la{\lambda}        \def\La{\Lambda}

         \def\Si{\Sigma}

\def\ps{\psi}
\def\om{\omega}

\def\d#1{\,\partial_{{#1}}}

\def\eom{equations of motion}

\def\vir{Virasoro}


\def\half{\textstyle {1\over2}}

\def\t#1{\dot{#1}}

\def\intsx{\int_\Si\int dx\,}
\def\intsxth{\int_\Si\int dx\,d\bar\th\,d\th\,}
\def\intx{\int dx\,}
\def\w{$W$-algebra}
\def\km{affine Lie}

\def\cs{Chern-Simons}
\def\3{three dimension}
\def\wi{Ward identities}
\def\cft{conformal field theory}

\def\mc{Maurer-Cartan equation}
\def\2{two dimension}
\def\paper{paper}

\def\H{$\cal H$}
\def\nfour{$N=4$ super $W_3$ algebra}
\def\sca{superconformal algebra}
\def\f#1/#2.{{\textstyle{#1\over#2}}}
\def\J{J_{ab}}
\def\F{F_{ab}}
\def\A{A_{ab}}
\def\diffS{\hbox{{\rm diff}}S^1}
\def\gf{Gel'fand-Fuchs cocycle}
\def\cc{\chi_a\chi_a}
\def\cov{{}_{{\rm cov}}}
\def\ccc{\cc''\vert_{\cov}}
\def\D{\bar D}
\def\sds{\hbox{$\,\in\hphantom{}\kern -.8em+$}}
\def\Lab{\bar\La}
\def\chb{\bar\chi}
\def\del{\partial}
\def\Vir{\hbox{{\sl Vir}}}
\def\X{\bar X}
\def\psb{\bar\psi}
\def\ao{\al_0}
\def\ai{\al_1}
\def\Xb{\bar X}
\def\*{,\ \allowbreak}
\def\i{{\textstyle\sqrt{-1}}}
\def\pc{\hbox{p.c.}}

\vskip 1in
\centerline{\bf A CLASSICAL $N=4$ SUPER W-ALGEBRA}

\vskip 1.5in
\centerline{R.E.C. Perret}
\smallskip
\count0=0

\vskip .5cm
\centerline{\it Department of Physics and Astronomy}
\centerline{\it University of Southern California}
\centerline{\it Los Angeles, CA 90089--0484}
\vskip 1in
\centerline{Submitted to: \it Int.\ J.\ Mod.\ Phys.\ A}

\vskip .5in

\beginsection{Abstract:}
I construct classical superextensions of the Virasoro algebra
by employing the Ward identities of a
linearly realized subalgebra. For the $N=4$ superconformal algebra,
this subalgebra is generated by the
$N=2$  $U(1)$ supercurrent and a spin~0 $N=2$ superfield.
I show that this structure can be extended to an $N=4$ super
$W_3$ algebra, and give the complete form of this algebra.
\vskip .5in\noindent
USC--92/023\hfill November 1992

\vskip.5in\eject

\def\Z{1}
\def\FL{2}
\def\GD{3}
\def\cft{4}
\def\toda{5}
\def\mm{6}
\def\ds{7}
\def\ws{8}
\def\Bakas{9}
\def\NY{10}
\def\LPRSW{11}
\def\Larry{12}
\def\Oi{13}
\def\KB{14}
\def\KS{15}
\def\PST{16}

\beginsection{1. Introduction}
Extensions of the \vir\ algebra  containing higher spin currents,
so called \w s [\Z,\FL], play a central role
in many areas of two dimensional physics.
They originally appeared as Hamiltonian structures on integrable
hierarchies [\GD], and since have been recognized as a symmetry
in for example conformal field theories [\cft], Toda theories [\toda]
and  \2al gravity [\mm,\ds].
Recently, there has been progress in
constructing $W$-algebraic analogues of string theory [\ws].

Unfortunately, the explicit form  of \w s is very complicated,
except in the simplest cases when only a few currents of low spin
are present, and as a consequence, relatively few \w s have
been constructed explicitly.
For example, an infinite series of algebras $W_n$ are known to
exist containing currents for all values of the spin from 2 to $n$
[\FL,\toda],
but only $W_3$ [\Z] and $W_4$ [\Bakas] have been constructed explicitly.
The $W_n$ all have $N=2$ superextensions [\NY], but again only for $n=3$
[\LPRSW,\Larry] and $n=4$ [\Oi] the explicit structure is known.
Although the \vir\ algebra has superextensions with arbitrarily high $N$
[\KB], for the $W_3$ algebra no extensions with
$N>2$ have been constructed previously.

Direct constructions of \w s are facilitated considerably by
recognizing the maximal linearly realized subalgebra \H\
with respect to which all remaining currents transform as primaries,
and imposing
covariance with respect to this \H.
In general, \H\ is a proper extension of the \vir\ algebra.
In the case of $N=2$ supersymmetric \w s, \H\
is the $N=2$ superconformal algebra, and the construction can be
simplified considerably by using $N=2$ superfields.
In this \paper, I construct an $N=4$ super $W_3$ algebra.
In this case, \H\ is a proper subalgebra of the linear
$N=4$ \sca\ [\KS], generated by the $N=2$ supercurrent $J$
and a spin 0  $N=2$ superfield $\Phi$.
The structure of $\Phi$ is somewhat unusual: its lowest component
is the inverse derivative of a spin 1 current.
No nonlocalities occur, because only (super)derivatives
of $\Phi$ enter the algebra.

I will use the dual formalism recently introduced in [\Oi].
In this approach, one constructs a three dimensional action,
analogous in  some respects to a \cs\ action,
from which the \wi\ of the algebra follow as equations of motion.
This formulation contains, besides the generating currents,
a set of dual one-forms. These one-forms satisfy \mc s which also
follow as \eom\ from the action. Combined with the \wi, the \mc s form
an integrable system which is equivalent to the Poisson bracket formulation
of the algebra. The integrability of the system is equivalent to the
Jacobi identities of the Poisson bracket formulation.

The construction of the action bears a close resemblance
to the construction of correlators in conformal field theory,
in particular in the way the \wi\ of the subalgebra \H\ are utilized.
The spins in this subalgebra are descendants of the identity,
while the remaining spins are primaries with respect to \H.
The nontrivial task is  to construct a set of three point functions
for the primaries and their composites,
consistent with \H\ covariance.
All correlators involving \H\ currents then follow from the couplings of
the primaries by the \H\ \wi, and only a finite number of them are nonzero.
The three dimensional action is the effective action of all correlators.

The explicit form of the \H\ dressings of the primary couplings is extremely
complicated in all but the simplest cases, and they are responsible
for the unwieldy form of the largest \w s constructed thus far.
It is however very straightforward to determine these dressings.
They are simply the \H\ covariantizations of the terms
in the effective action determined  by the couplings of the primaries,
and can be constructed in the familiar Yang-Mills fashion
by exploiting the anomaly in the \H\ \wi.
This anomaly is analogous to the inhomogeneous terms in the
transformation law of Yang-Mills gauge fields, and can be used to cancel
terms in the \H\ variation of the effective action caused by derivatives.

The organization of the rest of this \paper\ is as follows:
In Section~2, I construct as an example the $SO(N)$ symmetric $N$-extended
\sca s and discuss the relationship between nonlinear and linear
versions of these algebras.
As a preparation to the  \nfour,
I discuss in Section~3 the $N=2$ superfield structure of the $N=4$ \sca.
In Section~4, I derive the couplings of the primaries of the \nfour,
and Section~5 contains some conclusions.
The full structure of the action is given in an appendix.

\beginsection{2. Linear and nonlinear \sca s}
In this paper I follow the dual approach
to \w s recently introduced in [\Oi].
To illustrate this formalism in a simple setting,
I first consider the generic nonlinear $SO(N)$ symmetric \sca\ [\KB].
The generators of this algebra are the stress tensor $T$, of spin 2,
$N$ supersymmetry generators $X_a$, of spin $\f3/2.$, and
a set of $SO(N)$ currents $J_{ab}$, of spin 1, where $a$, $b$ run
from 1 to $N$.
I also introduce a set of gauge potentials $\mu$, of spin $-1$,
$\chi_a$, of spin $-\half$, and $A_{ab}$, of spin 0.

Let the subalgebra \H\ described in the Introduction be
the semidirect sum of the \vir\ algebra and the
$SO(N)$ \km\ algebra.
The nontrivial primary fields are then the $X_a$, and there are no
couplings possible between these primaries. Thus, the only nontrivial
couplings are the \H\ dressings of the propagators,
and the \3al action takes the following simple form
$$S=\intsx(2TF-2X_a\nabla\chi_a+\J\F+\mu\mu'''+\ccc
	   +\half\A\A')\,.\eqno(2.1)$$
Here, $F=d\mu+\mu\mu'$ is the field strength of $\diffS$,
$\nabla\chi_a=d\chi_a-\half\mu'\chi_a+\mu\chi_a'+Q\A\chi_b$
is the covariant derivative of $\chi_a$, and
$\F=d\A+\mu\A'+QA_{ac}A_{cb}$ is the $SO(N)$ field strength.
Primes denote derivatives with respect to the $S^1$ coordinate $x$,
and $\chi''_a\vert_{\cov}$ denotes a
covariantized second derivative of $\chi_a$
which is given explicitly below.
All fields are forms on the \2al surface $\Si$ with values in the
tensor algebra of $\diffS$, and
the lagrangian  is a spin 1 two-form on $\Si$
(modulo an inhomogeneous term in its transformation law
which is the conformal anomaly).

The action (2.1) is a generalization of the \cs\ action for a
finite dimensional Lie group and contains in fact the $SO(N)$
\cs\ action. It consists of a kinetic piece pairing the
generators of the algebra and the gauge fields, and a
potential describing the various couplings between the fields.
The quadratic piece of the potential is diagonal and
corresponds to  the central extension of the algebra,
while all higher couplings are contained in the
covariantizations of $\ccc$.

Ignoring for the moment the matter fields
$X_a$, $\chi_a$, the equations of motion corresponding to (2.1) are
$$F=0\,,\eqno(2.2)$$
$$\F=0\,,\eqno(2.3)$$
$$\nabla T+\mu'''+\half\J\A'=0\,,\eqno(2.4)$$
$$\nabla\J+\A'=0\,.\eqno(2.5)$$
Eqs.~(2.2,3) are the \mc s of $\diffS\sds\hbox{Map}(S^1,SO(N))$
and eqs.\ (2.4,5)
are the \wi\ of the central extension ${\cal H}=\Vir\sds\widehat{SO(N)}$.
Using (2.2,3), one easily shows that the terms $\intx\mu\mu'''$
and $\half\intx\A\A'$ appearing in (2.1) are closed two forms.
They correspond to the \gf\ and the central extension of affine $SO(N)$
respectively.

The explicit form of $\chi''_a\vert_{\cov}$ is determined by
demanding that the remaining piece of the lagrangian $\intx\ccc$
is likewise closed, still using the decoupled \wi\ (2.4,5).
This is done in much the same way as in Yang-Mills
theory, exploiting the nonhomogeneity of (2.4,5)
to covariantize the derivatives in $\cc''$.
First, one calculates the exterior derivate of $\cc''$
and observes that it can be cancelled by adding terms linear in the
\H\ currents. One then calculates the exterior derivative of these additional
terms and cancels it by terms quadratic in the \H\ currents. At this point
the process must terminate since all \H\ currents have dimension $\ge1$,
and in general the process will terminate as long as all \H\ currents
have positive dimension.

In this way one obtains
$$\eqalign{\intx\ccc&=\intx(\cc''+\half\cc T
			   -2Q\chi_a\chi_b'\J\cr
		    &\qquad-\f1/8.\chi^2\J\J+Q^2\chi_a\chi_bJ_{ac}J_{cb})
		    \,.\cr}\eqno(2.6)$$
The $SO(N)$ coupling is determined by imposing the integrability condition
$d^2=0$ on the \eom\ of the full action (2.1). This gives $Q^2=-\f1/4.$.

Notice that for $N=2$ the terms quadratic in $J$ cancel. In this case,
the algebra is in fact linear, and the action can be rewritten
concisely in terms of superfields.

Let $\th=\th_1+\th_2$ and $\bar\th=\th_1-\th_2$ denote the chiral
combinations of two fermionic coordinates $\th_1$, $\th_2$,
and let $D=\d{\bar\th}+\th\d{x}$, $\D=\d{\th}+\bar\th\d{x}$
denote the corresponding superderivatives.
Then $\phi'=\half(\D D+D\D)\phi$ is the $x$-derivative,
while I define $\dot\phi=\half(\D D-D\D)\phi$.
Let $\mu$ and $J$ be $N=2$ superfields of spin $-1$
and 1 respectively, and consider the following action
$$S=\intsxth(2JF+\mu\dot\mu')\,,\eqno(2.7)$$
where $F=d\mu+\mu\mu'+\half\D\mu D\mu$.
The \eom\ corresponding to (2.7)
$$F=0\,,\eqno(2.8)$$
$$\nabla J+\dot\mu'\eqno(2.9)$$
are easily seen to be integrable.
Here, $\nabla J$ is the $N=2$ supercovariant derivative,
which for a field $\phi$ of spin $s$ and charge $q$ reads
$$\nabla\phi=d\phi+s\mu'\phi+\mu\phi'+\half\D\mu D\phi+\half D\mu\D\phi
	     +q\dot\mu\phi\,.\eqno(2.10)$$
In $x$-space, (2.7) reduces to the original action (2.1) for $N=2$.

For $N=3$, a linear algebra [\KS] can be constructed in the same way using
$N=3$ superfields. The action is
$$S=\intsx d^3\th\,(2\La F+\mu D^3\mu)\,,\eqno(2.11)$$
where $\La$ is a spin $\half$ supercurrent,
$F=d\mu+\mu\mu'+\f1/4.D_a\mu D_a\mu$ and $D^3=D_1D_2D_3$.
One easily verifies that the \eom\ are integrable
so that this action indeed defines a linear $N=3$ \sca.
The $x$-space action is
$$\eqalign{S=\intsx&(2TF-2X_a\nabla\chi_a+2J_aF_a-2\La\nabla\la\cr
	    &\quad+\mu\mu'''+\cc''+A_aA_a'+\la^2\cr
	    &\quad+\half\chi^2T+i\ep_{abc}\chi_a\chi_b'J_c
	     +\la\chi_aJ_a-\La\chi_aA'_a)\,.\cr}\eqno(2.12)$$
The difference with (2.1) is that the terms quadratic in $J$ have been
replaced by nonminimal couplings
involving the spin $\half$ fields $\La$, $\la$.
The \H\ subalgebra
is now the \vir\ algebra,
so that $X_a$, $J_a$ and $\La$ are all to be regarded as matter primaries.
The last two terms in  (2.12) represent the three-point coupling
of the primaries. The coefficients of these couplings
are obtained by imposing the integrability of the \eom.

For $N\ge4$, no simple superfield formalism is available since
in the present approach the
\gf\ does not have an extension to $N\ge4$ superspace.
For $N=4$, a linear algebra does exist [\KS]
the most general form of which has been given in [\PST].
The action for the (untwisted) singly extended algebra of [\KS] is
$$\eqalign{S=\intsx&(2TF-2X_a\nabla\chi_a+\J\F-2\La_a\nabla\la_a
		    +2K\nabla B\cr
	   &\quad+\mu\mu'''+\cc''+\half\A\A'+\la^2+BB'\cr
	   &\quad+\half\chi^2T-2Q\chi_a\chi'_b\J\cr
	   &\quad+\la_a\chi_b\tilde\J-\La_a\chi_b\tilde A'_{ab}
		+\la_a\chi_aK-\La_a\chi_aB')\,,\cr}\eqno(2.13)$$
where the tilde denotes dualization of the $SO(4)$ indices.
The $J^2$ terms of (2.1)
have now been replaced by terms involving
the $\La_a$ of spin $\half$ and the $U(1)$ current $K$,
with dual one-forms $\la_a$ and $B$  respectively.
In this case, the \sca\ already has three independent couplings.
These couplings are in general much harder to determine than the \H\
covariantizations, and it would be extremely hard to construct
a $W_3$ extension in this $x$-space formulation.
As I show in the next section, the action (2.13) can be written in terms of
$N=2$ superfields, and in this formulation  the entire potential is again
just a sum of covariantized central extensions, with no nontrivial couplings.
This formulation will be the basis of my construction of $N=4$ $W_3$.

\beginsection{3. $N=2$ superfields for the $N=4$ \sca}
The spin content of the $N=4$ \sca\ of the previous section is
$(2)^1\*(3/2)^4\*(1)^7\*(1/2)^4$, while the spins of the
corresponding one-forms are
$(-1)^1\*(-1/2)^4\*(0)^7\*(1/2)^4$.
These spins can be grouped into $N=2$ superfields as follows.
For the one-forms, one defines a spin $-1$ superfield $\mu$,
two spin $-\half$ superfields $\chi$, $\chb$ and a spin 0
superfield $A$, where the highest component of $A$ is the derivative
of the $U(1)$ connection $B$,
i.e.~$A\mapsto A+\th\la_-+\bar\th\la_++\th\bar\th B'$.
For the generators, one can then define a spin 1 superfield $J$
(the $N=2$ supercurrent of the previous section),
two spin $\half$ superfields $\La$, $\Lab$ and a spin 0
superfield $\Phi$, which reads explicitly
$\Phi\mapsto\del^{-1}K+\th\La_-+\bar\th\La_++\th\bar\th J$,
where $\del^{-1}K$ denotes the inverse $x$-derivative of the $U(1)$
current $K$,
and $J$ is one of the $SO(4)$ currents.
The action will be local as long as $\Phi$ occurs only through its
(super)derivatives or multiplied by a factor the highest component
of which is an $x$-derivative, as in the kinetic term
$\Phi\nabla A$.

The fermionic superfields can have a $U(1)$ charge $q$,
which turns out to be 0, and a super $U(1)$ charge $Q$,
which is the coupling to the super $U(1)$ connection $A$.
The $N=2$ \sca\ has a charge
conjugation symmetry, which I will refer to as ''parity''.
$J$ and $\mu$ have negative and positive parity respectively, and in general
bosonic generators and dual one-forms have opposite parity.
$\La$ and $\Lab$ are parity conjugates, as are $\chi$ and $\chb$,
and it turns out that $\Phi$ and $A$ have positive and negative
parity respectively.
The entire lagrangian has negative parity, and spin 0.

To write a central term for the spin 0 field $A$,
one again has to use the inverse derivative.
It is easy to check that the term $A\del^{-1}\dot A$ reduces
in $x$-space to the
usual central terms for the component fields.

The action for the $N=4$ \sca\ can now be written simply as
$$\eqalign{S=\intsxth&(2JF+\Lab\nabla\chi-\La\nabla\chb+2\Phi\nabla A
	     +\mu\dot\mu'+\chb\dot\chi\vert_{\cov}+A\del^{-1}\dot A)
	     \,,}
	     \eqno(3.1)$$
where $\dot\chi\vert_{\cov}$ denotes the derivative covariantized
with respect to the \H\ subgroup generated by $J$ and $\Phi$.
The \H\  \wi\ are
$$\nabla J+\dot\mu'+\half\D AD\Phi+\half DA\D\Phi=0\,,\eqno(3.2)$$
$$\nabla\Phi+\del^{-1}\dot A=0\,,\eqno(3.3)$$
and one easily obtains
$$\intx\chb\dot\chi\vert_{\cov}=
  \intx(\half\chb\chi T+Q(\chb\D\chi D\Phi+\chb D\chi\D\Phi
  -\chb\chi\Phi'))\,,\eqno(3.4)$$
where the charge is determined to be $Q^2=\f1/4.$.
Notice that in this expression $\Phi$ only enters  through its
(super)derivatives,
so that indeed no nonlocalities occur.
The fact that $\Phi$ enters the \wi\ (3.2,3) only through its derivatives
guarantees that this will be true in general.

In $x$-space, the action (3.1) expands to the action (2.13)
of the previous section.

For comparison with the $W_3$ algebra discussed in the next section,
I now discuss what happens  if one omits the superfields
$\Phi$, $A$.
In that case, the integrability condition $d^2=0$ fails on $\chi$,
and instead one has
$$d^2\chi=-\f1/8.(\chb'\chi-\chb\chi'+\D\chb D\chi+D\chb\D\chi)\chi\,.
  \eqno(3.5)$$
The terms between brackets are exactly the appropriate coupling
of a spin 0, parity-odd superfield, and this is in fact how one deduces
the parity of $A$.

If one tries to construct an $N=4$ super $W_3$ algebra with the
spin content expected for linearly realized supersymmetry,
i.e.~$(3)^1\*(5/2)^4\*(2)^6\*(3/2)^4\*(1)^1$, one encounters an exact
analogue of eq.~(3.5): $d^2$ vanishes on the spin $\f3/2.$
superfields up to a term which can be cancelled by coupling
to a parity-odd spin 0 superconnection.
Thus, the entire \H\ subalgebra of the $N=4$ \sca, generated by
$J$ and $\Phi$, carries over to the $W_3$ algebra.

\beginsection{4. The $N=4$ super $W_3$ algebra}
In this section, I show that a consistent set of couplings
can be constructed for the primaries of the $N=4$ super $W_3$ algebra.
I start with the canonical spin content of the last paragraph,
and introduce the spin 0 superfield $\Phi$ when such becomes
necessary.
Thus, consider the $N=2$ superfields $J$, of spin 1 and negative parity,
parity conjugates $X$, $\X$, of spin $\f3/2.$ and charge $\pm q$,
and  $S$, of spin 2 and positive parity.
The corresponding dual one-forms are the superfields $\mu$, of
spin $-1$ and positive parity,  parity conjugates $\psi$, $\psb$, of spin
$-\f3/2.$ and charge $\pm q$, and $\nu$, of spin $-2$ and negative parity.
At this point, the \H\  subgroup is just the $N=2$ \sca.

The following is a systematic way to
determine the couplings between the primaries and their composites.

{\sl i]:} Write down general ans\"atze $d\nu_i\mapsto\sum\om_{ij}$
for the derivatives of all primary one-forms $\nu_i$
in terms of the primary one-forms and currents. Here, $j$ labels
different sectors, the terms in $\om_{ij}$ only differing in their
derivative structure and not in their field content.

{\sl ii]:} First, consider only the coupling of each primary to the
\H\ subalgebra. Thus, set the \H\ covariant derivative of each primary
to zero, and calculate $d\om_{ij}$ for each $i$, $j$.
The terms containing derivatives of the anomaly in the \H\ \wi\
can be cancelled by higher order terms, but all other terms should
vanish.
(For example, in the present case where \H\ is the $N=2$ \sca, all terms which
do not contain $\dot\mu'$ or a derivative thereof should vanish.)
This condition determines most, if not all, relative coefficients
in each sector.

{\sl iii]} Next, turn on the anomalies. That is, for each current $S_i$
set $\nabla S_i+{\cal D}\nu_i=0$, where $\nu_i$ is the one-form dual to $S_i$
and the differential operator ${\cal D}$ is the propagator for $\nu_i$
determined by the central extension.
Each coupling $d\nu_i\mapsto\om_{ij}$ in the ansatz comes from a term
$S_i\om_{ij}$ in the potential, and gives rise to a term
${\cal D}\nu_i\om_{ij}$
in the derivative of the potential. Requiring the sum of these contributions
to vanish gives relations between the coefficients in different sectors.
This condition follows from the associativity of the operator product
algebra.

{\sl iv]} The final step also requires the knowledge of the rules
$d\nu_0\mapsto\om_{0j}$ where $\nu_0$ denote the one-forms belonging to \H.
These follow from the first order covariantizations of the central terms.
Then, calculate $d^2\nu_i$ using the full set of rules
$d\nu_i\mapsto\sum\om_{ij}$, and require these expressions to vanish.
This determines all remaining parameters in the ansatz.

In the present case, the action has the form
$$\eqalign{S&=\intsxth(2JF+2\Xb\nabla\ps-2X\nabla\psb+2S\nabla\nu\cr
  &\qquad+\mu\t{\mu}'+\psb\t{\ps}''-\f q/s.\psb\ps'''+\nu\t{\nu}'''
   +\hbox{higher couplings})\,,\cr}\eqno(4.1)$$
where $q$ and $s=-\f3/2.$ are the (unknown) charge and spin of $\ps$.
{}From the first order covariantizations  of the central terms one obtains
$$\eqalignno{d\mu\mapsto&\f1/3.(q^2-\f9/4.)\psb\ps''-\f1/3.(q^2-3)\psb'\ps'
		     +\f1/3.(q^2-\f9/4.)\psb''\ps
    +\f1/6.q\psb'\t{\ps}-\f1/6.q\t{\psb}\ps'-\f1/4.\t{\psb}\t{\ps}\cr
   &+\f1/3.(q+\f3/2.)\D\psb D\ps'-\f1/3.(q+\f3/2.)\D\psb'D\ps
    -\f1/3.(q-\f3/2.)D\psb\D\ps'+\f1/3.(q-\f3/2.)D\psb'\D\ps\cr
   &-2\nu\nu'''+3\nu'\nu''-\t{\nu}\t{\nu}'
    -\f3/2.D\nu\D\nu''+2D\nu'\D\nu'-\f3/2.D\nu''\D\nu\,.&(4.2)\cr
\noalign{\medskip
	 \hbox{After step {\sl ii]} the ans\"atze are the following}
	 \smallskip}
d\ps\mapsto&\ao\ps''\nu+\ai\t{\ps}'\nu\cr
  &+\f1/3.(q\ao-(q^2-\half)\ai)\psi'\t{\nu}+(-\ao+\half q\ai)\ps'\nu'
   +\f1/6.(\ao-q\ai)\t{\ps}\t{\nu}-\f1/4.\t{\ps}\nu'\cr
  &+(\half\ao-\f1/3.q\ai)\ps\nu''
   +\f1/3.(-q\ao+(q^2-\f3/4.)\ai)\ps\t{\nu}'\cr
  &+\half(\ao-(q+\half)\ai)D\ps'\D\nu+\half(\ao-(q-\half)\ai)\D\ps'D\nu
    &(4.3)\cr
  &+\f1/3.(-\ao+(q+\half)\ai)D\ps\D\nu'
   +\f1/3.(-\ao+(q-\half)\ai)\D\ps D\nu'\,,\cr
&\cr
d\nu\mapsto&\be_0(\f1/5.\nu\t{\nu}'+\f1/10.\nu'\t{\nu}
    +\f1/10.\D\nu D\nu-\f1/10.\D\nu'D\nu)&(4.4)\cr
   &+\ga_0(\psb\ps'-\psb'\ps)+\ga_1(\psb\t{\ps}-\t{\psb}\ps)
    +\ga_2D\psb\D\ps+\ga_3\D\psb D\ps\,,\cr}$$
modulo terms involving the zero-forms.
The latter follow from the terms in the potential of degree $\ge2$
in the zero-forms, which can be parametrized as
$$\eqalign{V_{\ge2}&=\ze_0\nu\nu'\Xb X+\ze_1(\nu\D\nu\Xb DX-\pc)
		     +\ze_2(\nu\D\nu D\Xb X-\pc)\cr
		   &\qquad+\eta_0\psb\ps\Xb X+\th_0(\nu\ps S\Xb-\pc)\,,\cr}
		   \eqno(4.5)$$
where $\pc$ denotes the parity conjugate.
The $\ga_i$ and $\ze_i$ are restricted by the relations
$-\half\ga_0+(1+q)\ga_1-(\f3/2.-q)\ga_3=0$,
$-\half\ga_0-(1-q)\ga_1-(\f3/2.+q)\ga_2=0$,
$\half\ze_0-\ze_1-4\ze_2=0$.

Step {\sl iii]} determines the $\ga_i$ in terms of the $\al_i$ and $q$
$$\eqalignno{\ga_0&=-\f1/9.(q^2-\f9/4.)(\ao-q\ai)\,,&(4.6)\cr
	     \ga_1&=\f1/18.(q^2-\f9/4.)\ai\,,&(4.7)\cr
	     \ga_2&=\f1/18.(q-\f3/2.)(\ao-\ai)\,,&(4.8)\cr
	     \ga_3&=-\f1/18.(q+\f3/2.)(\ao+\ai)\,,&(4.9)\cr}$$
consistently with the above conditions on the $\ga_i$.
The charge is determined in step {\sl iv]} to be $q=-\f5/2.$, up to a sign
which corresponds to interchanging $\ps$ and $\psb$.
Continuing with step {\sl iv]} one finds
$$\halign to \hsize
{\qquad#\hfil&\qquad#\hfil&\qquad#\hfil&
 \qquad#\hfil\tabskip=0pt plus 10cm&\hfil#\tabskip=0pt\cr
 $\ao=\f3/2.\i$,&$\ai=-\f3/2.\i$,&&&\cr
 $\be_0=5\i$,&&&&\cr
 $\ga_0=\i$,&$\ga_1=-\half\i$,&$\ga_2=-\i$,&$\ga_3=0$,&\cr
 $\ze_0=-12$,&$\ze_1=-6$,&$\ze_2=0$,&&(4.10)\cr
 $\eta_0=12$,&&&&\cr
 $\th_0=12$,&&&&\cr}$$
where the sign ambiguity of $\i$ corresponds to the freedom of flipping
the signs of $\nu$ and $S$.

With the above choice for the parameters, $d^2$ vanishes on all one-forms
except on $\ps$, where instead one finds
$$\eqalign{d^2\ps=(&-3(\psb\ps'''-\psb'\ps''+\psb''\ps'-\psb'''\ps
		       +\t{\psb}\t{\ps}'-\t{\psb}'\t{\ps})\cr
		   &+5(\t{\psb}\ps''-\t{\ps}'\ps'+\t{\psb''}\ps
		       +\psb\t{\ps}''-\psb'\t{\ps}'+\psb''\t{\ps})\cr
		   &-2(\D\psb D\ps''-\D\psb'D\ps'+\D\psb''D\ps)\cr
		   &+8(D\psb\D\ps''-D\psb'\D\ps'+D\psb''\D\ps))\psi\,.\cr}
  \eqno(4.11)$$
Notice that the expression multiplying $\ps$ has odd parity.
Also, its highest component is a total $x$-derivative, and therefore it
can potentially be cancelled by coupling $\psi$ to the spin 0 superfield
$A$ of the $N=4$ \sca,
so that its covariant derivative becomes
$$\nabla\ps=d\ps-\f3/2.\mu'\ps+\mu\ps'-\f5/2.\t{\mu}+QA\ps\,.\eqno(4.12)$$
One finds that indeed the right hand
side of (4.11) is cancelled by the first order covariantizations of
the central terms for $\psb$, $\ps$ if the coupling constant
$Q=\sqrt{-6}$.

This completes the analysis of the elementary couplings between the
primaries. To construct the full action is now a matter of covariantizing
all terms in the lagrangian implied by these couplings.
Although the result, which is given in an appendix,
is extremely lengthy, this is a
straightforward computation, and could presumably have been done
by a machine.

\beginsection{5. Conclusion}
In this paper I gave some examples of classical \w s in a dual formulation,
and in particular constructed a new $N=4$ super extension of
the $W_3$ algebra.
I illustrated how the explicit structure of \w s can be split into a set
of couplings between generators which are primaries of
a subalgebra \H, and covariantizations with respect to this subalgebra.
In practice, the couplings between the primaries are usually found to
be much simpler in structure than the \H\ covariantizations, which can be
extremely lengthy.
I showed that the couplings can be constructed independently
of the bulk of the \H\ covariantizations,
and although I have not proven this, the existence
of a consistent set of couplings presumably
implies the existence of the whole algebra.

\beginsection{Acknowledgements:}
I would like to thank P.~Bouwknegt and K.~Pilch for reading
the manuscript.
This research was partially
supported by the Department of Energy Contract \#DE--FG03--91ER40168.

\def\ps{\psi}
\def\X{\bar X}
\def\psb{\bar\psi}
\def\t#1{\dot#1}

\def\D{\bar D}
\def\5#1{#1^{(5)}}
\def\4#1{#1^{(4)}}

\def\D{\bar D}
\def\f#1/#2.{{\textstyle{#1\over#2}}}
\def\hlf{\f1/2.}

\def\P{\Phi}
\def\dP{\dot\Phi}
\def\dJ{\dot J}
\def\dS{\dot S}

\def\dr{\dot\r}
\def\ds{\dot \s}

\def\th{\r\s'''}
\def\tw{\r\s''}
\def\on{\r\s'}
\def\z{\r\s}

\def\dth{\r\ds'''}
\def\dtw{\r\ds''}
\def\don{\r\ds'}
\def\dz{\r\ds}

\def\Btw{\r\D\s''}
\def\Bon{\r\D\s'}
\def\Bz{\r\D\s}

\def\Dtw{\r D\s''}
\def\Don{\r D\s'}
\def\Dz{\r D\s}
\def\DD{\D\P D\P}
\def\DPD{\D\P'D\P}
\def\pd{(\P'-\dP)}
\def\pc{\hbox{p.c.}}

\beginsection{Appendix}
In this appendix I give the complete lagrangian for the
$N=4$ super $W_3$ algebra
of Section~4 in $N=2$ superspace.
Below, $Q=\sqrt{-6}$ denotes the charge of the fermionic superfields.

$$\eqalignno{\cal L=&\hskip.7mm
   JF+\X\nabla\ps-X\nabla\psb+S\nabla\nu+\Phi\nabla A\cr
  &+\hlf A\partial^{-1}\t{A}+\hlf\mu\t{\mu}'\cr
  &+\hlf\psb\t{\ps}''-\f5/6.\psb\ps'''\cr
  \global\def\r{\psb}\global\def\s{\ps}
  &-\f11/2.\tw J+\f5/2.\dtw J+\hlf\Don\D J-2\Bon DJ\cr
  &-\f19/4.\on J'-\f5/12.\on\dJ+\f5/4.\dz J'+\f1/4.\dz\dJ\cr
  &+\f1/3.\Dz\D J'-\f4/3.\Bz DJ'-\f4/3.\z J''\cr
  &+Q(\hlf\Btw D\P+\hlf\Dtw\D P+\f5/2.\tw\dP-\hlf\tw\P'\cr
  &+\Bon D\P'+\Don\D\P'+\f5/2.\on\dP'-\on\P''-\hlf\dz\dP'\cr
  &+\hlf\Bz D\P''+\hlf\Dz\D\P''+\f5/6.\z\dP''-\hlf\z\P''')\cr
  &-\f35/3.\on J^2+3\dz J^2-\f14/3.\Bz JDJ+\f4/3.\Dz J\D J\cr
  &-\f2/3.\z J\dJ-10\z JJ'+\f4/3.\z\D JDJ\cr
  &+Q(\f5/2.\Bon JD\P+\f5/2.\Don J\D\P\cr
  &+11\on J\dP-\f5/2.\on J\P'-\hlf\on\D JD\P-2\on DJ\D\P-\f5/2.\dz J\dP\cr
  &+\f5/2.\Bz JD\P'+\f5/2.\Dz J\D\P'+2\Bz DJ\dP-\hlf\Dz\D J\dP\cr
  &+\f5/4.\Bz J'D\P+\f5/4.\Dz J'\D\P+\f1/4.\Bz\dJ D\P+\f1/4.\Dz\dJ\D\P\cr
  &+\f11/2.\z J\dP'-\f5/2.\z J\P''-\hlf\z\D JD\P'-2\z DJ\D\P'\cr
  &+\f19/4.\z J'\dP+\f5/12.\z\dJ\dP-\f5/4.\z J'\P'-\f1/4.\z\dJ\P'\cr
  &-\f1/3.\z\D J'D\P-\f4/3.\z DJ'\D\P)\cr
  &-\f11/4.\tw\DD+\f5/4.\don\DD\cr
  &-\Bon D\P\P'+\Bon D\P\dP-\f1/4.\Don\D\P\P'-\f1/4.\Don\D\P\dP\cr
  &-\f31/12.\on\DPD-\f13/6.\on\DD'-\f5/24.\on\P'^2
   +\f5/24.\on\dP^2\cr
  &+\f3/4.\dz\DPD+\hlf\dz\DD'+\f1/8.\dz\P'^2-\f1/8.\dz\dP^2\cr
  &-\f4/3.\Bz(D\P'\P'+D\P\P''-D\P'\dP-D\P\dP')\cr
  &-\f1/3.\Dz(\D\P'\P'+\D\P\P''+\D\P'\dP+\D\P\dP')\cr
  &-\f4/3.\z(\D\P''D\P+2\DPD'+\DD'')\cr
  &+Q^2(-\hlf\tw\DD-\Bon\dP D\P-\Don\dP \D\P\cr
  &-\on\DPD-\on\D\P  D\P'-\f5/2.\on\dP^2+\on\dP\P'+\hlf\dz\dP^2\cr
  &-\hlf\Bz\dP'D\P-\hlf\Dz\dP'\D\P-\Bz\dP D\P'-\Dz\dP\D\P'\cr
  &-\hlf\z\D\P''D\P-\z\DPD'-\hlf\z\DD''\cr
  &-\f5/2.\z\dP'\dP+\hlf\z\dP'\P'+\z\P''\dP)\cr
  &-8\on J^3\cr
  &+Q(3\Bz J^2D\P+3\Dz J^2\D\P\cr
  &-3\z J^2\P'+\f35/3.\z J^2\dP-\f4/3.\z J\D JD\P-\f14/3.\z JDJ\D\P)\cr
  &+3\dz J\DD-\f35/3.\on J\DD\cr
  &-\f7/3.\Bz(DJ\DD-J\dP D\P+J\P'D\P)\cr
  &+\f2/3.\Dz(\D J\DD-J\dP \D\P-J\P'\D\P)\cr
  &-\f1/3.\z\dJ\DD-5\z J'\DD
   +\f2/3.\z\D j\pd D\P+\f2/3.\z DJ(\P'+\dP)\D\P\cr
  &-\f16/3.\z J\DPD-\f14/3.\z  J\DD'-\f1/3.\z J\P'^2+\f1/3.\z J\dP^2\cr
  &+Q^2(-\f5/2.\on J\DD-\f5/2.\Bz J\dP D\P-\f5/2.\Dz\dP\D\P\cr
  &-\f5/2.\z J\DPD-\f5/2.\z J\DD'+\f5/2.\z J\dP\P'
   -\f11/2.\z J\dP^2\cr
  &+\hlf\z\D J\dP D\P+2\z DJ\dP\D\P-\f5/4.\z J'\DD-\f1/4.\z\dJ\DD)\cr
  &+Q(\f19/4.\on\dP\DD-\f5/4.\dz\dP\DD\cr
  &+\f3/4.\Bz D\P'\DD+\hlf\Dz\D\P'\DD-\f9/8.\Bz\dP^2D\P+\f1/8.\Dz\dP^2\D\P\cr
  &+\Bz\dP\P'D\P+\f1/4.\Dz\dP\P'\D\P+\f1/8.\Bz\P'^2D\P+\f1/8.\Dz\P'^2\D\P\cr
  &+\f9/4.\z\dP'\DD'+\f21/12.\z\dP\DPD+\f5/12.\z\P'\DPD
   +\f7/4.\z\dP\DD\cr
  &-\f5/24.\z\dP^3+\f5/24.\z\dP\P'^2+\f1/8.\z\P'\dP^2-\f1/8.\z\P'^3)\cr
  &+Q^3(\on\dP\DD+\hlf\Bz\dP^2D\P+\hlf\Dz\dP^2\D\P\cr
  &+\hlf\z\dP'\DD+\z \dP\DPD+\z\dP\DD'-\hlf\z\dP^2\P'+\f5/6.\z\dP^3)\cr
  &-(12+3Q^2)\z J^2\DD\cr
  &+(10Q+\f5/2.Q^3)\z J\dP\DD\cr
  &+(\hlf-\f15/8.Q^2-\hlf Q^4)\z\dP^2\DD+(-\hlf-\f1/8.Q^2)\z\P'^2\DD\cr
  \global\def\r{\nu}\global\def\s{\nu}
  &+\hlf\dth\cr
  &+\th J+3\Btw DJ+\don\dJ+4\Bon DJ'+\hlf\dz\dJ'+\f3/2.\Bz DJ''\cr
  &-\hlf\don J^2-\Bz JDJ-\hlf\dz JJ'+2\on J\dJ-\f15/4.\on\D JDJ\cr
  &-\hlf\Bz JDJ'+\f9/4.\Bz\dJ DJ-\f3/4.\Bz J'DJ\cr
  &+\hlf\th\DD+\f3/2.\Btw(P'-\dP)D\P\cr
  &+\hlf\don\DPD-\hlf\don\DD'+\hlf\don\P'^2-\hlf\don\dP^2\cr
  &+2\Bon(\P''-\dP')D\P+2\Bon\pd D\P'\cr
  &+\f1/4.\don\D\P''D\P-\f1/4.\don\DD''+\hlf\P'\P''-\hlf\don\dP\dP'\cr
  &+\f3/4.\Bz(\P'''D\P+2\P''D\P'+\P'D\P''-\dP''D\P-2\dP'D\P'-\dP\P'')\cr
  &-\Bz J^2DJ-\on J^3\cr
  &-\hlf\don J\DD-\hlf\Bon J\pd D\P-\hlf\Bon DJ\DD\cr
  &-\f1/4.\dz J'\DD-\hlf\dz J\DD'\cr
  &+\on\dJ\DD-\f15/8.\on\D J\pd D\P\cr
  &-2\on J\DD'+\on J\P'^2-\on J\dP^2\cr
  &-\f1/4.\Bz DJ'\DD+\f9/8.\Bz\dJ\pd D\P-\f3/8.\Bz J'\pd D\P\cr
  &+\f3/4.\Bz\DPD-\f3/2.\Bz\DD'+\f9/8.\Bz\P'^2-\f9/8.\Bz\dP^2\cr
  &-\f1/4.\Bz J(\P''D\P+\P'D\P'-\dP'D\P\dP D\P')\cr
  &-\f3/2.\on J^2\DD-\Bz JDJ\DD-\hlf\Bz J^2\pd D\P\cr
  &-\f23/16.\on\dP^2\DD+\f23/16.\on\P'^2\DD\cr
  &+\f5/8.\Bz\pd D\P'\DD-\f9/16.\Bz\P'\dP^2D\P-\f9/16.\Bz\dP\P'^2D\P\cr
  &+\f9/16.\Bz\P'^3D\P+\f9/16.\Bz\dP^3D\P\cr
  \global\def\r{\psi}\global\def\s{\nu}
  &-2\alpha_0\X(\r''\s-\dr'\s-\hlf D\r'\D\s-\D\r'D\s\cr
  &+\f13/12.\r'\ds+\f1/4.\r'\s'-\f1/4.\dr\ds+\f1/4.\dr\s'\cr
  &+\f1/3.D\r\D\s'+\f2/3.\D\r D\s'_\f1/3.\r\s''-\r\ds'\cr
  &+\f17/3.\r'\s J-3\dr\s J-\f5/3.\D\r D\s J-\f5/3.D\r\D\s J\cr
  &+\f13/3.\D\r\s DJ-\f5/3.D\r\s\D J-\f11/3.\Bz DJ-\f7/3.\Dz\D J\cr
  &+\f7/3.\dz J-\on J-\f5/3.\z\dJ-\z J'\cr
  &+Q(-\D\r'\s D\P-D\r'\s\D\P-2\r'\s\dP+\r'\s\P'+\dr\s\dP\cr
  &-\D\r\s D\P'-D\r\s\D\P'+\r\s\P''-\r\s\dP'\cr
  &+\hlf\r'\D\s D\P-\r'D\s\D\P+\D\r D\s\dP+\hlf D\r\D\s\dP\cr
  &-\f1/4.\D\r\ds D\P-\f1/4.D\r\ds\D\P+\f1/4.\D\r\s'D\P+\f1/4.D\r\s'\D\P\cr
  &-\f13/12.\r\ds\dP-\f1/4.\r\s'\dP+\f1/4.\r\ds\P'-\f1/4.\r\s'\P'\cr
  &+\hlf\r\D\s D\P'-\r D\s\D\P'-\f1/3.\Bon D\P+\f2/3.\Don\D\P)\cr
  &+8\z J^2\cr
  &+Q(-3\D\r\s JD\P-3D\r\s J\D\P+\f5/3.\r\D\s JD\P-\f5/3.\Dz J\D\P\cr
  &-\f17/3.\z J\dP+3\z J\P'+\f13/3.\z DJ\D\P+\f5/3.\z\D JD\P)\cr
  &-\f4/3.\z\DPD+\f1/3.\z\DD'-\f5/6.\z\P'^2+\f5/6.\z\dP^2\cr
  &+Q^2(\r's\DD+\f1/4.\dz\DD-\f1/4.\on\DD\cr
  &+\D\r\s\dP D\P+D\r\s\dP\D\P-\hlf\Bz\dP D\P-\Dz\dP\D\P\cr
  &+\z\DPD+\z\DD'+\z\dP^2-\z\dP\P')\cr
  &+(8+3Q^2)\z J\DD\cr
  &-\f3/2.Q\z(\dP+\P')\DD-Q^3\z\dP\DD)\cr
  \global\def\r{\nu}\global\def\s{\nu}
  &+\beta_0(\hlf\don S+\hlf\Bon DS+\f1/4.\dz S'+\f3/20.\on\dS
   +\f3/10.\Bz DS'\cr
  &+\f7/10.\on JS+\f9/10.\Bz SDJ-\f1/10.\Bz JDS\cr
  &+\f7/20.\on\DD S+\f9/20.\Bz\pd D\P S-\f1/20.\Bz\DD DS)\cr
  \global\def\r{\psb}\global\def\s{\psi}
  &+\gamma_0(-\f4/3.\on S+\f2/3.\Bz DS-\z S'+\f1/3.\z\dS-4\z JS\cr
  &+Q(\f4/3.\z\dP S-\f2/3.\z\D\P  DS)-2\z\DD S)\cr
  \global\def\r{\nu}\global\def\s{\nu}
  &+\zeta_0(\hlf\on\X X-\hlf\Bz\X DX-\hlf Q\Bz D\P\X X)\cr
  &+\hlf\eta_0\psb\psi\X X+\theta_0\nu\psb SX\cr
  &-\pc}$$


\def\eef #1 {\llap{#1\enspace}\ignorespaces}
\def\ref{\beginsection{References}\frenchspacing
    \parindent=0pt\leftskip=1truecm\parskip=8pt plus 3pt
    \everypar={\eef}}
    \def\npb#1{Nucl. Phys. {\bf B#1}}
    \def\cmp#1{Commun. Math. Phys. {\bf#1}}

    \def\plb#1{Phys. Lett. {\bf #1B}}
    \def\ijmp#1{Int. J. Mod. Phys. {\bf A#1}}

    \def\mpl#1{Mod. Phys. Lett. {\bf A#1}}
    \def\faa#1{Funct. Anal. Appl. {\bf#1}}
    
    \def\tmf#1{Teor. Mat. Fiz {\bf#1}}

\def\hb{\hfil\break}

\ref

[1] A.B. Zamolodchikov, \tmf{65} (1985) 347

[2] V.A. Fateev and A.B. Zamolodchikov, \npb{280 [FS18]} (1987) 644;
\hb V.A. Fateev and S.L. Luk'yanov, \ijmp{3} (1988) 507

[3] I.M. Gel'fand and L.A. Dickey, Russ. Math. Surv. {\bf 30} (1975) 77;
    \faa{10} (1976) 4; \faa{11} (1977) 93

[4] V.A. Fateev and A.B. Zamolodchikov, \tmf{71} (1987) 163;
\hb F. Bais, P. Bouwknegt, M. Surridge and K. Schoutens, \npb{304} (1988)
    348, 371

[5] A. Bilal and J.-L. Gervais, \npb{314} (1989) 646; \npb{318} (1989) 579

[6] J. Goeree, \npb{358} (1991) 737;
\hb M. Fukama, H. Kawai and R. Nakayama, \cmp{143} (1992) 371

[7] I.R. Klebanov and A.M. Polyakov, \mpl{6} (1991) 3273;
\hb E. Witten, \npb{373} (1992) 187

[8] S. Das, A. Dhar and S. Rama, \mpl{6} (1991) 3055; \ijmp{7} (1992) 2295;
\hb C.N. Pope, L.J. Romans and K.S. Stelle, \plb{269} (1991) 287;
\hb C.N. Pope, L.J. Romans, E. Sezgin and K.S. Stelle, \plb{274} (1992) 298;
\hb M, Berschadsky, W. Lerche, D. Nemeschansky and N. Warner,
    CERN--TH--6582--92;
\hb C.N. Pope, E. Sezgin, K.S. Stelle and X.J. Wang, CTP TAMU--64/92

[9] I. Bakas, \plb{213} (1988) 313;
\hb R. Blumenhagen, M. Flohr, A. Kliem, W. Nahm, A. Recknagel, R.
    Varnhagen \npb{361} (1991) 255

[10] D. Nemeschansky and S. Yankielowicz, USC--91/005A

[11] H. Lu, C.N. Pope, L.J. Romans, X. Shen and X.-J. Wang,
     \plb{264} (1991) 91

[12] L.J. Romans, \npb{369} (1992) 403

[13] R.E.C. Perret, USC--92/016 ({\tt hepth/9208068})

[14] V.G. Knizhnik, \tmf{66} (1986) 68;

[15] M. Adamollo, L. Brink, A. D'Adda, R. D'Auria, E. Napolitano, S. Sciuto,
     E. Del Giudice, P. Di Vecchia, S. Ferrara, F. Gliozzi, R. Musto and
     R. Pettorino, \plb{62} (1976) 105; \npb{114} (1976) 297;
\hb K. Schoutens, \npb{295} (1988) 634
\hb M. Berschadsky, \plb{174} (1986) 285

[16] A. Sevrin, W. Troost and A. van Proeyen, \plb{208} (1988) 447

\vfil\eject\end